# GHz operation of a quantum point contact using stub-impedance matching circuit

Anusha Shanmugam [a], Prasanta Kumbhakar [a], Harikrishnan Sundaresan [a],

Annu Anns Sunny [a], J L Reno [b], and Madhu Thalakulam [a, *]

[a]*Indian Institute of Science Education & Research Thiruvananthapuram, Kerala 695551, India*
[b]*Center for Integrated Nanotechnologies, Sandia National Laboratory, Albuquerque, NM, United States of America*

## Abstract

Quantum point contacts (QPC) are the building blocks of quantum dot qubits and semiconducting quantum electrical metrology circuits. QPCs also make highly sensitive electrical amplifiers with the potential to operate in the quantum-limited regime. Though the inherent operational bandwidth of QPCs can eclipse the THz regime, the impedance mismatch with the external circuitry limits the operation to a few kHz regimes. Lumped-element impedance-matching circuits are successful only up to a few hundreds of MHz. QPCs are characterised by a complex impedance consisting of quantized resistance, capacitance, and inductance elements. Characterising the complex admittance at higher frequencies and understanding the coupling of QPC to other circuit elements and electromagnetic environments will provide valuable insight into its sensing and backaction properties. In this work, we couple a QPC galvanically to a superconducting stub tuner impedance matching circuit realised in a coplanar waveguide architecture to enhance the operation frequency into the GHz regime and investigate the electrical amplification and complex admittance characteristics. The device, operating at ~$1.96 \, GHz$ exhibits a conductance sensitivity of $2.92 \times 10^{-5} (e^2/h)/\sqrt{Hz}$ and a bandwidth of $13 \, MHz$. Besides, the RF reflected power unambiguously reveals the complex admittance characteristics of the QPC, shining more light on the behaviour of quantum tunnel junctions at higher operational frequencies.

# 1. Introduction

Devices exhibiting quantum tunnelling are the main building blocks of quantum technology. Superconducting tunnel junctions, the Josephson junctions, are the major circuit components of a superconducting qubit and Josephson voltage standards[1–3], while semiconducting qubits [4–8] and charge counting devices are formed by the quantum point contacts (QPC), a semiconducting tunnelling device. QPCs are voltage-controlled constrictions defined in a two-dimensional electron gas (2DEG) by a pair of gates[9]. The transport through the constriction is ballistic and is quantized in $2e^2/h$, when the voltage controlling the width of the constriction is varied [10] as explained by the Landauer-Büttiker formalism [11]. Once the conductance falls below $G_0 = 2e^2/h$, the transport enters the pure tunnelling regime, and all the QPC-based devices are operated in this regime. In the tunnelling regime, the QPC conductance is highly sensitive to changes in the electrical environment, and by virtue of this, they make excellent electrical amplifiers with the potential to reach the quantum-limited detection regime[12]. They are regarded as the readout devices for semiconductor quantum dot qubits[13,14] and also find applications in other sensing technologies, such as displacement and strain [15,16]. Though the inherent bandwidth of QPCs falls in the THz regime, the impedance mismatch with the low-impedance measurement circuitry limits the operational speed of these devices to the kHz regime in the conventional operational scheme [17,18]. Radio-frequency reflectometry techniques exploiting lumped element impedance matching circuits have pushed the operational frequency into a few hundreds of MHz[19]. When it comes to high-frequency measurements, QPC is no longer treated as a simple resistor; one needs to consider the reactive components too [20,21]. It has been shown theoretically that not only the resistance but also the capacitive and inductive contributions show discrete behaviour [20–24]. There are a few reports on the discrete behaviour in the imaginary part of admittance as well [25–28]. In this context, the change in quantum capacitance as a result of the change in dot occupation is exploited in the dispersive gate sensing technique [29].

It is very important to achieve GHz frequency operation since the charge readout and error correction protocols in qubit circuits demand highly sensitive, rapid, and non-invasive techniques, at least a few orders faster than the coherence times [30,31]. Apart from this, the important energy scales, such as single-triplet splitting, interdot tunnel coupling, and Zeeman splitting in the quantum dots, fall in the GHz frequency range[32,33]. The reactive elements of the QPC may have a decisive role in its interaction with the associated devices and environment, which can shed more light on aspects such as measurement noise and backaction.

The parasitic capacitance of the matching circuit limits the operating frequency of radio-frequency QPCs to the sub-GHz regime, which also puts a limitation on the maximum load that can be transformed to 50 $\Omega$[34,35]. The stub-matching technique derived from microwave impedance matching enables operation at much higher operating frequencies, yields a larger bandwidth, and has a tunable load resistance[36–39]. A stub tuner is a pair of terminated transmission lines; their electrical length determines the operating frequency, while their difference determines the matched load [36]. For a single-stub shunt tuning circuit, the length of transmission line is chosen such that the admittance of the combined line and load is of the form $Y = \frac{1}{Z_0} + iB$, where $Z_0$ is the characteristic impedance of the transmission line and $B$ is the susceptance. Shunting the line at length $D_1$ with an open transmission line of length $D_2$ with an admittance $-iB$ transforms the impedance of the load to 50 $\Omega$.

The input impedance of the circuit is given by the parallel combination of the two arms,

$$Z_{total} = Z_0 \left[ tanh(\gamma D_2) + \frac{Z_0 + Z_L tanh(\gamma D_1)}{Z_L + Z_0 tanh(\gamma D_1)} \right]$$

Where $\gamma = \alpha + i\beta$, is the propogation constant. For the lossless case, the attenuation constant $\alpha = 0$. $\beta$ is the phase constant of the transmission line ($\beta = 2\pi f \sqrt{\varepsilon}/c$), where $f$ is the frequency. Then the reflection coefficient is given by $\Gamma = \frac{Z-Z_0}{Z+Z_0}$. The lengths $D_1$ and $D_2$ are chosen in such a way that the reflection coefficient vanishes at the resonant frequency $f$.

An in-situ tunable shunt stub matching circuit for QPCs has been previously demonstrated [40]. High-frequency characterization of a gold break-junction[37] and carbon nanotubes[39] has also been carried out using stub-impedance matching circuits. The use of a stub-matching circuit for high-frequency charge sensing operations has not yet been demonstrated. Being a planar architecture, the stub matching circuit possess a great potential for scalability.

In this work, we demonstrate the radio frequency operation of a GaAs/AlGaAs QPC electrical amplifier using a stub tuner based impedance matching circuit. The stub tuner made of Aluminium in a coplanar waveguide geometry was designed to impedance match the QPC operating point $\sim \frac{e^2}{h} \sim 25 \, k\Omega$. The experiment is carried out utilizing an RF reflectivity setup [12] at a frequency $\sim 1.96 \, GHz$, the resonant frequency of the system. We determine the sensitivity of the QPC by applying a small excitation signal to the QPC gate, modulating the conductance and the reflected power from the stub tuner. From the signal-to-noise ratio (SNR) of the resulting sidebands in the reflected signal, we obtain a conductance sensitivity of $2.92 \times 10^{-5} (e^2/h)/\sqrt{Hz}$ with a detection bandwidth of $13 \, MHz$. Apart from the high-frequency operation, we also probe the complex impedance of the QPC at the resonant frequency which is not accessible by conventional transport measurements. We observe plateaus in the low-frequency pinch-off characteristics of the QPC's corresponding quantized conductance values. We also observe discrete changes in the magnitude and phase of the high-frequency reflected signal ($S_{11}$) against the QPC gate voltage in congruence with the quantized conductance steps, a manifestation of quantized capacitance and inductance in the QPC.

## 2. Device and measurement

All the measurements are done in a cryogen-free dilution refrigerator with a base temperature of 10 mK. A schematic representation of the measurement circuit is shown in Fig. 1(a). The circuit supports both low-noise DC and high-frequency measurements. The high-frequency measurements are conducted exploiting an RF reflectometry circuit, as shown in Fig. 1(a)[12]. The reflected RF power at resonance from the QPC-stub tuner system is recorded as a function of various operating parameters. This will help one observe the transport characteristics of the QPC at frequencies and band widths prescribed by the stub tuner. We use a bias tee to apply the RF and DC signals to the QPC simultaneously. High-frequency signals after passing through the QPC is grounded to the stub tuner ground plane via a capacitor, while the DC or low-frequency part of the current is returned to room temperature to perform a direct measurement of the current through the device. DC and low-frequency source-drain voltages are applied using a low-noise source measure unit (SMU) and lock-in amplifier, respectively. The DC voltage and the small-signal AC excitations are applied onto the QPC gate electrode using a passive room-temperature AC+DC adder circuit, the gate box.

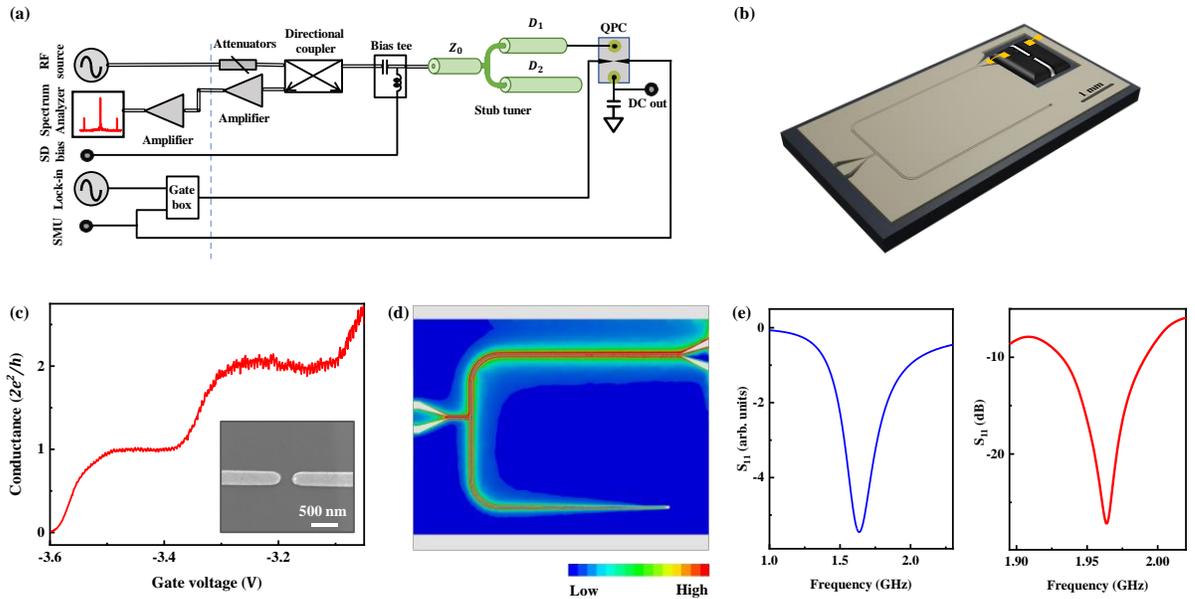

Fig. 1. (a) Schematic drawing of the RF reflectometry setup in dilution refrigerator (b) Optical image of the stub tuner similar to the one measured with a schematic of QPC showing the coupling scheme. (c) Pinch-off characteristics of the QPC at the base temperature. Inset: SEM image of the QPC (d) Ansys HFSS simulation showing the electric field distribution of the stub tuner at resonance. (e) Left: Simulated $S_{11}$ (blue). Right: $S_{11}$ of the stub tuner when the QPC is pinched off at base temperature (red).

The device consists of an Aluminium stub tuner galvanically coupled to a GaAs/AlGaAs QPC, as shown in Fig. 1(b). The stub tuner is fabricated in a coplanar wave guide (CPW) configuration on a $\sim 500\ \mu m$ thick sapphire substrate using photolithography, followed by a $\sim 150\ nm$ thick Aluminium metallization. The width of the stub tuner center conductor is $\sim 50\ \mu m$ and the gap between the center conductor and ground plane $\sim 29\ \mu m$ yielding a characteristic impedance of $50\ \Omega$ for the CPW. The open and shunted arm lengths of the stub tuner, $D_1$ and $D_2$ are $0.972\ cm$ and $1.028\ cm$ respectively. The QPC is fabricated on a GaAs/AlGaAs wafer hosting a 2DEG with a carrier concentration of $\sim 2.02 \times 10^{11} cm^{-2}$ and a mobility of $\sim 2.13 \times 10^{6} cm^2 V^{-1} s^{-1}$. The surface gates forming the QPC are defined using a combination of photolithography and electron-beam lithography followed by Cr/Au metallization, while the ohmic contacts to the 2DEG are achieved by the indium alloying technique. Fig. 1(c) shows the transconductance of the QPC, and the inset to Fig. 1(c) shows a scanning electron microscope (SEM) image of the QPC. The transconductance shows clear quantized conductance plateaus and an eventual pinch-off of the 1D channel for gate voltages $\lesssim -3.6\ V$.

We conduct a three-dimensional high-frequency simulation of the stub tuner/QPC system, assuming a $25\ k\Omega$ resistor in place of the QPC, using ANSYS HFSS. The simulation yields a resonant frequency of $1.65\ GHz$. The electric field distribution for the combined system at resonance $\sim 1.65\ GHz$, shown in Fig. 1(d) confirms that the QPC is located at the electric-field maximum of the stub tuner, which is required for a good coupling between the stub tuner and the QPC. The left panel in Fig. 1(e) shows the simulated two port reflectivity (S$_{11}$) data, yielding a resonant frequency of $\sim 1.65\ GHz$. The S$_{11}$ data, measured using a vector network analyzer for the stub tuner/QPC system with the entire measurement circuit at a temperature of $10\ mK$, yields a resonant frequency of $\sim 1.963\ GHz$ as shown in the right panel of Fig. 1(e). We believe that the difference in the simulated and measured $S_{11}$ characteristics is due to the parasitic impedances introduced by the normal metal connecting wires, ohmic contacts, and the 2DEG source and drain regions on either side of the QPC, which were not accounted for in the simulation.

## 3. Results

Now we explore the complex admittance characteristics of the QPC from the RF reflectivity of the stub tuner. Within the limit of appreciable coupling between the QPC and the stub tuner, the resonance characteristics of the latter should be influenced by the operating conditions of the former. Fig. 2(a) shows a plot of the magnitude of $S_{11}$ in the vicinity of the resonance as the QPC conductance is varied starting from the pinched-off regime through the first conductance plateau to the second conductance plateau. The inset shows a magnified view of the center of the resonance, showing a progressive change in the reflected power at resonance. The complex admittance of a QPC includes the resistive component at the constriction and the reactive components, viz., the inductive and capacitive contributions. The variation in the phase of the reflected power should carry information on the contributions from the change in the reactive components as the channel dimensions and the population of the modes are varied. Fig. 2(b) shows a plot of the change in the phase of the reflected power as the conductance is varied from the pinched-off regime through the second plateau. We assume the phase is zero when the QPC is completely pinched-off so that there is no tunnel junction present [41]. The inset shows a magnified view of the change in phase in the vicinity of the resonant frequency as the conductance is varied.

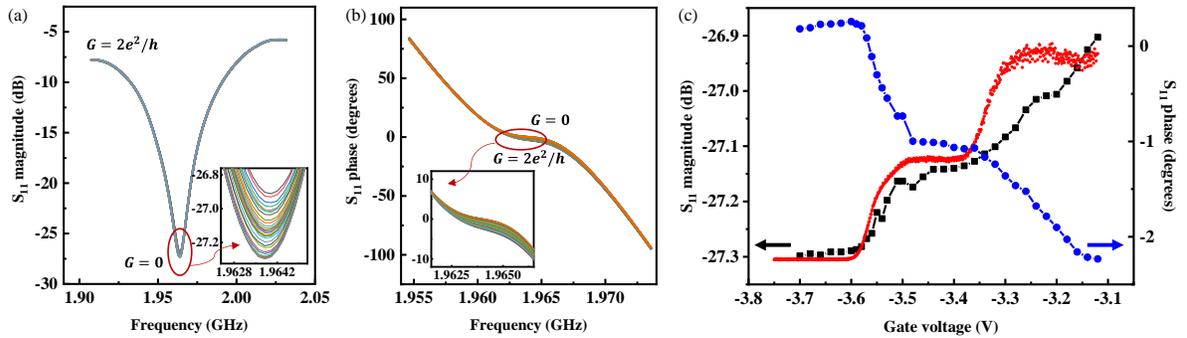

Fig. 2. (a) Magnitude and (b) phase of $S_{11}$ measured for different gate voltages at base temperature. Magnified views in the vicinity of the center of the resonance are given in the respective insets. (c) Magnitude (blue squares) and phase (black circles) of $S_{11}$ plotted as a function of the QPC gate voltage. Transconductance of the QPC showing conductance plateaus (red scattered dots) are plotted as a guide to the eye.

The change in the magnitude and phase of the reflected power against the gate voltage controlling the QPC conductance is shown in Fig. 2(c). The QPC conductance showing quantized conductance plateaus is also shown in red as a guide to the eye. We find that both the magnitude and phase of $S_{11}$ also exhibit a step-like change similar to that of the conductance. The two-terminal QPC can be considered a self-inductance connected in series with a parallel connection comprising a capacitor and a resistor[23]. The quantum inductance is determined by the harmonic mean of the velocities of the propagating electron modes, while the quantum capacitance is specified by the reflecting modes. The admittance of a QPC can be expressed as $G(\omega) = G_0 - i\omega E$, where $G_0$ is the conductance and $E$ is the emittance[20]. It has been shown that, with the opening of channels, the capacitance and the emittance decrease in a steplike manner in synchronism with the conductance steps[22]. As we increase the conductance by increasing the gate voltage from the pinched-off state, the device enters the pure tunnelling regime. Below the first plateau, the QPC reactance consists of only resistive and capacitive elements. On the first plateau, the source and drain sides are connected by a 1D mode of conductance $2e^2/h$, which contributes to the inductance. Calculations on the admittance of QPC using the Wigner function description formalism has also shown stepwise behaviour in both the real and imaginary parts of admittance[23]. Whenever the conductance is constant, the number of open transmission modes is unchanged, and the phase of $S_{11}$ is also unchanged. We also observe a similar behaviour at 1.6 K and 4 K (data not shown).

From our observations so far, we find that the complex reflectance of the stub tuner is highly influenced by the conductance of the QPC. The performance of the QPC as an electrical amplifier depends on the coupling between the QPC and the stub tuner and, in turn, the former's capability to modulate the reflectance of the latter. To calibrate the sensitivity of the QPC, we examine the SNR of the sideband in the reflected power at resonance resulting from small-signal gate excitations. For this, we bias the QPC along the steeper region of the transconductance trace such that the conductance of the channel is $\sim 0.5\ G_0$ and apply a 10 $kHz$ gate exaction of amplitude 1 $mV_{RMS}$ (0.0024 $e^2/h$) onto one of the QPC gates while an input RF power $P_{rf} \sim 51\ dBm$ at resonance (1.96 $GHz$) is applied at the input port of the sub-tuner. The amplitude-modulated stub tuner reflected power, amplified by both the cryogenic and the room-temperature high-frequency amplifiers, as shown in Fig. 3(a), is acquired using an RF spectrum analyzer. The conductance variation due to the gate excitation amplitude modulates the reflected signal, producing side bands at frequencies $v_c - v_m$ and $v_c + v_m$. From the SNR of the side-bands, we can calculate the conductance sensitivity of the device $\delta G$ using the formula[34], $\delta G = \frac{1}{2} \frac{\Delta G}{\sqrt{BW} 10^{SNR/20}} = 2.92 \times 10^{-5} (e^2/h)/\sqrt{Hz}$, where $BW$ is the measurement bandwidth of the external circuit, set by the measurement bandwidth of the spectrum analyzer in our case. Major operating parameters influencing the sensitivity of the QPC are the operating point defined by the gate voltage, $P_{rf}$, and the gate excitation. Now we inspect and optimise each of these to identify the best operating condition of the sensor.

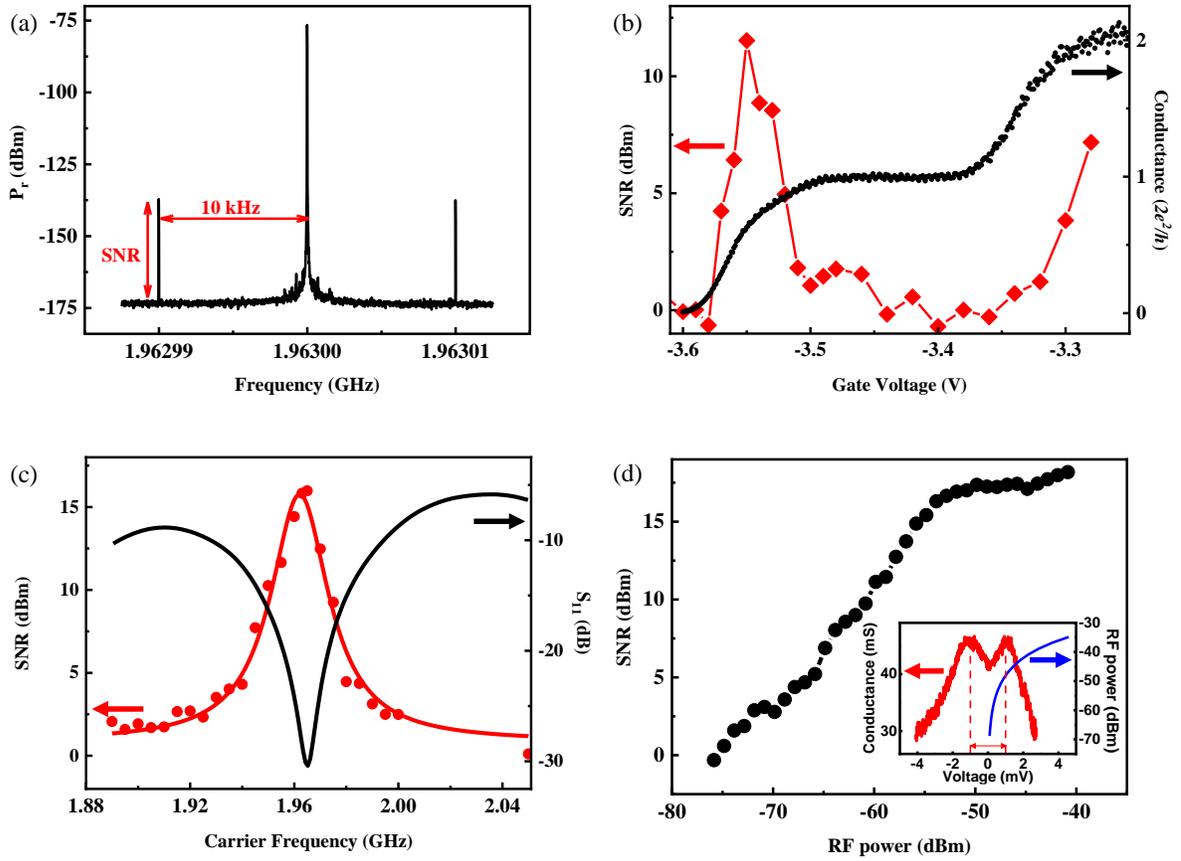

Fig. 3. (a) Amplitude modulated reflected signal ($P_r$) showing side bands with 10 kHz, $\sim 0.024\ e^2/h$ gate excitation. (b) SNR vs gate voltage and pinch-off characteristics of the QPC plotted together (c) SNR vs carrier frequency across the resonance showing the detection bandwidth. A plot of $S_{11}$ showing the bandwidth is mostly dictated by the stub tuner resonance characteristics (d) SNR vs input RF power. Inset: Conductance vs source-drain bias (red, left) and RF power (blue, right) the calculated RF voltage on the sample.

For optimising the QPC conductance, we apply a $10\ kHz$ gate excitation of amplitude $1\ mV_{RMS}$ from the lock-in amplifier while $P_{rf} \sim 51\ dBm$ is applied at the input port of the stub tuner. Further, the SNR of the side bands of the amplitude-modulated reflected power is acquired using the spectrum analyzer for various QPC conductance values by varying the QPC gate voltage. Fig. 3(b) shows the SNR vs. the gate voltage controlling the QPC conductance. The pinch-off characteristics of the QPC is also shown in black as a guide to the eye. It can be seen from the plot that the SNR is high in the nonlinear region of the QPC transconductance and goes to a minimum when the channel is pinched-off and also on the first conductance plateau. The maximum SNR corresponds to the point where the channel conductance $G \sim \frac{e^2}{h}$, mid-way between the pinch-off and the first conductance plateau. To find the detection bandwidth of the system, we study the SNR as a function of carrier frequency by applying $P_{rf} \sim 51\ dBm$, with a $10\ kHz$ gate excitation of amplitude at $1\ mV_{RMS}$, while maintaining the QPC conductance $\sim \frac{e^2}{h}$ ($V_g = -3.56\ V$). Fig. 3(c) shows a plot of the SNR against the carrier frequency (left); the RF reflectivity $S_{11}$ is also shown (right) for comparison. From a Lorentzian fit to the data, we extract a bandwidth of 13 $MHz$, the frequency span by which the SNR falls by 3 $dB$ of the peak value, which is consistent with the bandwidth of the resonance at 1.96 $GHz$.

Fig. 3(d) shows the SNR vs. $P_{rf}$ at the resonant frequency 1.96 $GHz$, while the QPC was biased close to $G \sim \frac{e^2}{h}$ with a small signal 1 $mV_{RMS}$, 10 $kHz$ excitation on the gate. We see that for a clear two-decade change in the power, the SNR assumes a linear relationship with the power, as expected in the case of amplitude modulation. For $P_{rf}$ beyond $\sim -50\ dBm$ the SNR slowly saturates. To understand this, we inspect the span of the RF voltage along the source-drain voltage axis for various $P_{rf}$ and compare it with the QPC conductance as a function of source-drain voltage, as shown in the lower-right inset to Fig. 3(d). The red trace (left axis) corresponds to the QPC conductance, while the blue trace (right axis) maps $P_{rf}$ against the resulting RF voltage swing on the QPC. The RF voltage is calculated by taking into account the power gain at the resonance due to the Q-factor ~ 258, and the voltage coupling factor $\lambda_g = \sqrt{\pi Z_0 / R} = 0.016$, where $R$ is the two-probe resistance of the sample at the QPC operating regime [12]. We see a close correspondence between the I-V characteristics and the saturation of charge sensitivity at higher RF input powers. The SNR increases as long as the RF voltage is confined to the increasing part of the conductance, while it starts saturating once the RF voltage starts sampling regions in source-drain bias where the conductance starts decreasing. The saturation of SNR happens for RF voltages beyond $\sim 2\ mV$ across the sample which is represented by the vertical dashed line in the inset to Fig 3(d). This also confirms the stub tuner's capability to probe the transport characteristics of the QPC in a frequency regime not accessible by conventional transport measurements.

## 4. Conclusion

In this work, we have integrated a superconducting stub-impedance matching circuit to enhance the operational frequency of a QPC electrical amplifier into the GHz regime. The QPC, operating at a frequency of $\sim 1.96\, GHz$, exhibits a conductance sensitivity of $2.92 \times 10^{-5} (e^2/h)/\sqrt{Hz}$ with a band width of $13\, MHz$. Tunnel junctions possess complex admittance characteristics and are described by a combination of capacitance, resistance, and inductance. Besides enhancing the operation bandwidth of the QPC amplifier, the stub tuner also helps to probe the complex admittance characteristics and understand the behaviour of the reactive elements at very high frequencies. From the measured complex reflection coefficient of the stub tuner, we capture not only the change in the resistance of the QPC but also the variation in the capacitance and inductance values as the channel width and admittance are varied. QPC, characterised by a discrete number of one-dimensional modes, exhibits discrete capacitance and inductance changes as the number of modes is altered. Our attempt to understand the transport characteristics is an important exercise since the inter-dot tunnel coupling and Zeeman splitting in quantum dot circuits are usually in the GHz range[33,42]. The reactive elements, such as the inductances and capacitances, play crucial roles in the QPCs coupling to high-frequency electromagnetic modes in the environment. Understanding the behaviour of the reactive elements can shed more light on deeper aspects of quantum sensing such as backaction due to the QPC's noise on the measurement system.

**Author contributions**: MT conceived the problem. The GaAs/AlGaAs heterostructure wafer is grown by JLR. AS & PK fabricated the devices. AS, PK, HS & AAS conducted the measurements. AS analyzed the data and, AS & MT co-wrote the manuscript.

**Acknowledgements:** MT acknowledges the funding support received from DST, Govt. of India under the grand ST/ICPS/QuST/Theme-4/2019 and MoE, STARS under the grand MoE-STARS/STARS1/363. This work was performed, in part, at the Centre for Integrated Nanotechnologies, an Office of Science User Facility operated for the U.S. Department of Energy (DOE) Office of Science. Sandia National Laboratories is a multimission laboratory managed and operated by National Technology & Engineering Solutions of Sandia, LLC, a wholly owned subsidiary of Honeywell International, Inc., for the U.S. DOE's National Nuclear Security Administration under contract DE-NA-0003525. The views expressed in the article do not necessarily represent the views of the U.S. DOE or the United States Government.